\documentclass[runningheads]{llncs}
\usepackage[T1]{fontenc}
\usepackage{graphicx}
\usepackage{amssymb}
\usepackage{booktabs}
\usepackage{changepage}
\usepackage{hyperref}
\begin{document}
\title{Fantastic (small) Retrievers and How to Train Them: \textsc{mxbai-edge-colbert-v0} Tech Report}
\titlerunning{Mxbai-edge-ColBERTv0}

\makeatletter
\newcommand{\intern}{\textsuperscript{*}}   
\newcommand{\joint}{\textsuperscript{\dag}} 

\newcommand{\printtitlefootnotes}{%
  \begingroup
    \renewcommand{\thefootnote}{\fnsymbol{footnote}}%
    \footnotetext[1]{Work performed during an internship at Mixedbread.}
  \endgroup
}
\makeatother
\author{%
  Rikiya Takehi\inst{1,2}\intern \and
  Benjamin Clavié\inst{1} \and
  Sean Lee\inst{1} \and
  Aamir Shakir\inst{1}
}
\authorrunning{R. Takehi et al.}
%
\institute{Mixedbread AI \and
Waseda University \\
\email{\{rikiya,ben,sean\}@mixedbread.com}}
\maketitle              
\printtitlefootnotes
\setcounter{footnote}{0}
\begin{abstract}
In this work, we introduce mxbai-edge-colbert-v0 models, at two different parameter counts: 17M and 32M. As part of our research, we conduct numerous experiments to improve retrieval and late-interaction models, which we intend to distill into smaller models as proof-of-concepts. Our ultimate aim is to support retrieval at all scales, from large-scale retrieval which lives in the cloud to models that can run locally, on any device. mxbai-edge-colbert-v0 is a model that we hope will serve as a solid foundation backbone for all future experiments, representing the first version of a long series of small proof-of-concepts. As part of the development of mxbai-edge-colbert-v0, we conducted multiple ablation studies, of which we report the results.
In terms of downstream performance, mxbai-edge-colbert-v0 is a particularly capable small model, outperforming ColBERTv2 on common short-text benchmarks (BEIR) and representing a large step forward in long-context tasks, with unprecedented efficiency.
\end{abstract}
\section{Introduction}

In the last two years, neural Information Retrieval (IR) has experienced an unprecedented level of interest, owing in large part to the rapid development and deployment of Large Language Models (LLMs) and the proven effectiveness of Retrieval Augmented Generation (RAG) pipelines~\cite{rag}, where retrieval models are used to provide LLMs with useful context.

As part of this wave, end-user interest in multi-vector retrieval methods, also called late interaction models or, more simply, ColBERT, after the model which initially introduced this method~\cite{colbert}. Where the dominant paradigm in neural IR, Dense Passage Retrieval (DPR)~\cite{dpr}, leverages a single, large vector to represent documents, ColBERT models instead employ numerous smaller vectors, with each individual token representation projected to a small dimension then retained. In order to make this tractable, ColBERT models are frequently used with aggressive index quantization~\cite{colbertv2,plaid} or as second-stage rankers in a larger pipeline.

The growing popularity of multi-vector models can be explained by their retrieval performance. ColBERT models have been noted for their particularly robust out-of-domain performance~\cite{colbertv2}, especially in multi-modal settings~\cite{modernvlm}. They have also recently been demonstrated to provably alleviate certain limitations of single-vector retrieval approaches, with a 150M parameter ColBERT models vastly outperforming 8B parameter single-vector embeddings on benchmarks designed to test the limits of embedding models~\cite{LIMIT}.

In spite of these strong performances, the ecosystem for open ColBERT models have moved more slowly than that of single-vector models. Up until last year, the most widely used ColBERT model was ColBERTv2, originally released in 2021. Subsequently, answerai-colbert-small-v1\footnote{\url{https://huggingface.co/answerdotai/answerai-colbert-small-v1}} demonstrated a 33 million parameter ColBERT model could outperform all existing small retrievers, and reached performance exceeding even that of ColBERTv2 and most <500M parameter retrievers.

\begin{figure}
    \centering
     \includegraphics[width=0.95\linewidth]{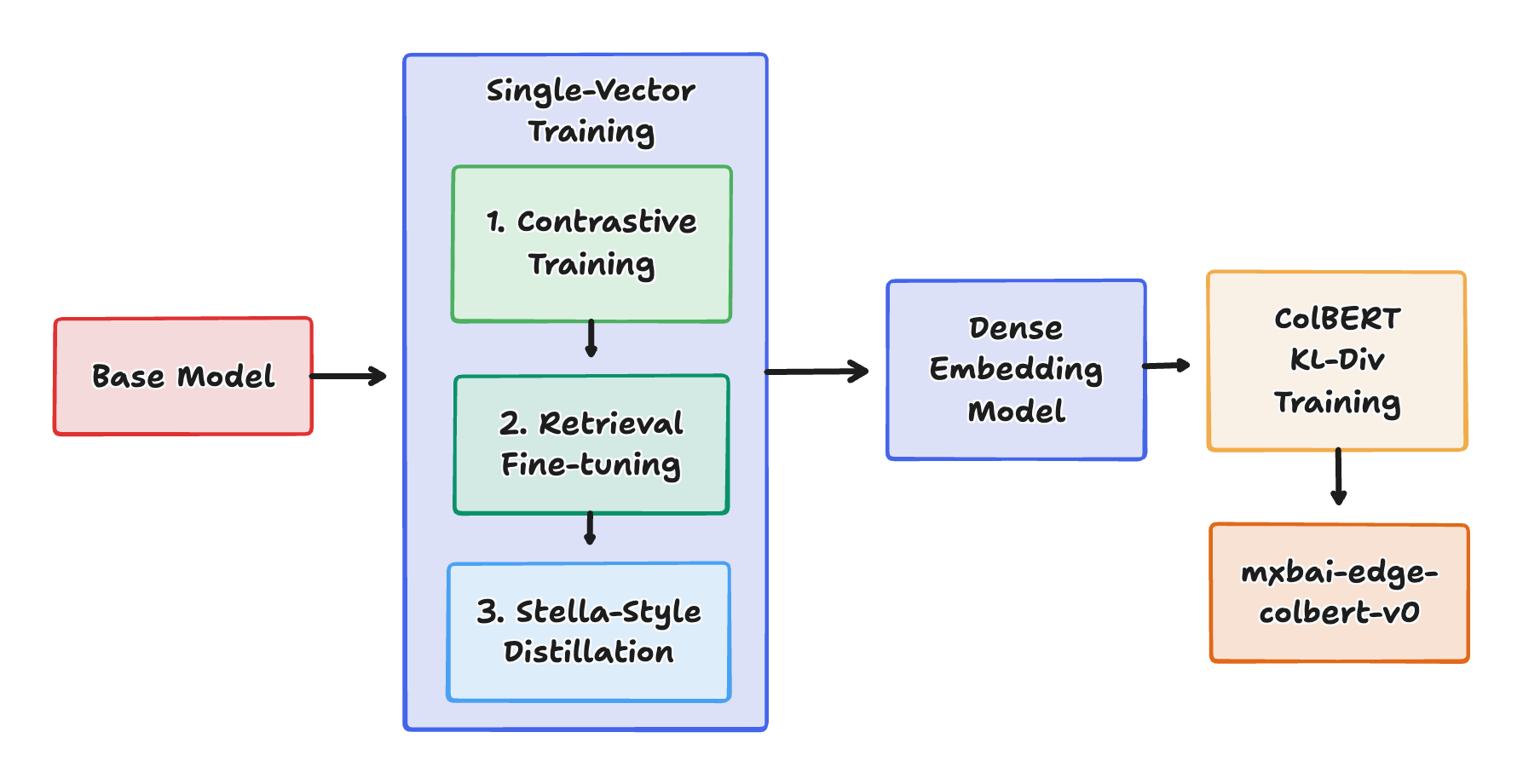}
    \caption{An overview of the full training process}
    \label{fig:overview}
\end{figure}

However, there has, until very recently, been a lack of late-interaction models featuring modern features, such as long context handling due to backbone limitations, at least in the text modality. Indeed, both ColBERTv2 and answerai-colbert were built on top of BERT variants, namely the original BERT~\cite{bert} and MiniLM~\cite{minilm}, with short context limits and poor efficiency, especially across longer contexts.

ModernBERT~\cite{modernbert} spearheaded a new wave of novel encoders, built with efficiency in mind and allowing long-context encoders. Following its original release, it has been followed by Ettin, a reproduction of it across model sizes, and ModernVBERT, which combines Ettin with a vision-encoder, bringing the architectural improvements to multi-modality. GTE-ModernColBERT\footnote{\url{lightonai/GTE-ModernColBERT-v1}}~\cite{gtemoderncolbert} was subsequently released, leveraging ModernBERT as a backbone, and creating the new de-facto standard for 130M+ parameter ColBERT, outperforming ColBERTv2 and all dense retrievers in its parameter class.

A large gap, however, remains: while GTE-ModernColBERT is a strong, ``full-sized'' model, answerai-colbert-small-v1 remains, by far, the most downloaded ColBERT model, in spite of its architectural limitations. In our own exploratory work at Mixedbread, we found ourselves frequently using it, as its small size and strong performance provided a strong testbed for various experiments. We firmly believe that performance at both ends of the scale spectrum is very important, especially as small models are strong predictor of the performance impact of model modifications.

As such, we decided to address this gap, and create the mxbai-edge-colbert-v0 family of ColBERT models. These models come in two different sizes, with 17 and 32 million parameters, and have been created to serve as a strong baseline to support further experiments while addressing the needs of users seeking a modern, low parameter-count ColBERT. To train these models, we first created dense embedding baselines through a series of three training stages, before running numerous ablations resulting in the released models.

The resulting models are strong performers across the board, with considerably improved efficiency over previous models. Notably, mxbai-edge-colbert-v0-17m outperforms ColBERTv2 despite an embedding dimension of 48, one third of the commonly used 128, and an extremely low compute and memory footprint. Its strong performance, combined with long-context handling and very low latencies, makes it particularly suitable for re-ranking applications on-device.

These models, the first of an hopefully long series of efficient edge models, represent a solid foundation for further studies on the effectiveness of ColBERT model. We hope that they will support research, both within and outside of Mixedbread, while supporting a large range of real-world uses.

\section{Creating a Suitable Dense Base Model}

Previous work has demonstrated the importance of beginning ColBERT training from a suitably ``warmed-up'' model, with considerably better results obtained when training from a dense embedding model rather than initializing training from scratch~\cite{answeraicolbert,gtemoderncolbert}, outperforming even those obtained by further training an existing ColBERT model~\cite{jacolbertv2.5,colbertv2}. We believe this effect to be due to the fact that dense embedding models now routinely undergo a long, distantly-supervised contrastive alignment training phase~\cite{e5} before being fine-tuned on high quality data, which is not commonly done for ColBERT models\footnote{As the aim of this work is to create a suitable backbone to identify the effect of individual modifications, we leave the exploration of a ColBERT-specific contrastive warm-up phase to future work, but believe it holds strong potential for further improvements.}.

In light of this, we first set out to create a suitable dense backbone, at our target model sizes: 32 and 17 million parameters. We use the Ettin~\cite{ettin} encoder models as starting models, which are a replication of the ModernBERT training recipe across various model sizes~\cite{modernbert}.

\subsection{Contrastive Pre-Training}

We follow the standardised recipe for our contrastive pre-training phase, as is now commonly adopted by the large majority of embedding models~\cite{e5,snowflake,gte}. Effectively, this phase consists in leveraging many open datasets during which we have approximate queries that can be mapped to documents that are at least somewhat semantically related to them. In practice, this takes many different forms: forum posts with their title acting as a query, QA pairs extracted from common websites, etc. This training is done with a large batch size, facilitated by GradCache~\cite{gradcache}, and resulting in a better embedding alignment.

For this section, we used the contrastor training framework, which was used in the training of the Nomic embeddings models~\cite{nomic}. We used a common selection of pre-training datasets, presented in Table~\ref{tab:datasets}. We follow the work done on mxbai-embedding-large~\cite{mxbaiembed} and train sequentially, that is, one dataset at a time, rather than all at once, which we empirically found to result in better performance. A similar form of this effect was described in the snowflake-arctic embeddings tech report~\cite{snowflake}, where stratification of training examples by origin dataset yielded superior results.

\begin{table}[h!]
\label{tab:datasets}
\centering
\caption{Datasets used for contrastive pretraining.}
\begin{tabular}{l r}
\hline
\textbf{Dataset} & \textbf{Size (rows)} \\
\hline
synthetic datasets & 2.65M \\
nomic-embed-unsupervised-data\footnote{Minus the reddit set, which we observed as detrimental to performance.} & 172.8M \\
bge-m3-data & 1.57M \\
cornstack (subsampled, 8\% total) & 20M \\
\hline
\textbf{Total} & 197M \\
\hline
\end{tabular}
\end{table}

Interestingly, this pre-training phase highlighted an effect that appears common to all the ModernBERT and Ettin models: a higher learning rate is needed to reach satisfying results when compared to previous backbone encoder models. This effect was first described in the ModernBERT paper, where hyperparameter sweeps revealed that a considerably higher learning rate was necessary for ModernBERT to outperform previous encoders on common retrieval tasks~\cite{modernbert}. Table~\ref{tab:hyperparams} shows the NanoBEIR NDCG@10 of multiple training runs with different learning rates.


\begin{table}[h!]
\centering
\caption{Performance of two models post-contrastive training with varying learning rates (NDCG@10 on NanoBEIR)}
\setlength{\tabcolsep}{12pt} 
\renewcommand{\arraystretch}{1.2}
\begin{tabular}{c c c c}
\hline
\textbf{Model} & \textbf{Learning Rate} & \textbf{Batch Size} & \textbf{NDCG@10} \\
\hline
17M & 3.5e-04 & 24576 & 0.493 \\
17M & 6.0e-04 & 24576 & \textbf{0.523} \\
\hline
32M & 2.8e-04 & 12288 & 0.543 \\
32M & 5.0e-04 & 12288 & \textbf{0.559} \\
\hline
\end{tabular}
\label{tab:hyperparams}
\end{table}

\subsection{Fine-tuning}
Subsequently, we move on to the next step of dense embedding pre-training: supervised fine-tuning on higher quality data, with mined hard negatives~\cite{minednegs}. The mining of hard negatives is a key factor in training embedding models, as it helps provide stronger ``counter-examples'': for every example that is relevant to the query, we also provide the models with examples that are not. When using solely random negatives, this task becomes trivial: a completely unrelated negative will rapidly only reach very low similarity scores, and stop meaningfully contributing to learning. This also dulls the learning process of what a ``match'' looks like: if the negative examples are always completely unrelated, then the model does not have to work as hard to learn what makes a document truly relevant. Hard negative mining attempts to solve this problem by gathering a set of harder negatives that look more similar to the positive document. 

This ensures that the model learns to accurately represent details that differentiate relatively similar documents, rather than just general topics. On the other hand, negatives that are too hard can also be harmful to the learning process: if the model only ever sees negative examples that are ``almost-positives'', then it might fail to learn good high-level representations. Moreover, negatives that are too hard carry a high false-negative rate: most datasets commonly used for retrieval have sparse labels, and it is highly likely that a lot of the highest-scoring negative documents could actually be positives.

As such, crafting a good mix of negatives is important. We follow NV-Embedv2~\cite{nvretriever} in our mining process, where we used Qwen3-Embedding-8B to mine hard negatives and set the threshold to 0.95. To learn on various negative hardness, we also mixed the data with 35\% of BM-25 mined and 30\% of randomly mined documents. We mine negative for the de-facto standard set of training datasets for retrieval fine-tuning: MSMARCO, NQ, HotPotQA and PubMed.

\begin{table}[h!]
\centering
\caption{Performance comparison of Mxbai Edge models (Dense) with and without finetuning (NDCG@10).}
\setlength{\tabcolsep}{14pt}
\renewcommand{\arraystretch}{1.2}
\begin{tabular}{l c}
\hline
\textbf{Model} & \textbf{NDCG@10} \\
\hline
Mxbai Edge 17M (Dense, non-FT) & 0.523 \\
Mxbai Edge 17M (Dense, FT)     & \textbf{0.556} \\
\hline
Mxbai Edge 32M (Dense, non-FT) & 0.559 \\
Mxbai Edge 32M (Dense, FT)     & \textbf{0.576} \\
\hline
\end{tabular}
\label{tab:mxbai_edge_ft}
\end{table}

We adopt the AnglE training loss~\cite{angle} for this fine-tuning step, using the AnglE codebase\footnote{\url{https://github.com/SeanLee97/AnglE}}. We train on a selection of common datasets, once again seeking to follow standard practices while avoiding over-contamination in relation to frequent benchmarks.

\subsection{``Stella-style'' Distillation}

Finally, we add a third stage to our model pre-training, which is inspired by the Stella~\cite{stella} model family. Stella, in addition to more commonplace retrieval training, introduces \textbf{embedding space distillation}: it generates embeddings for queries and documents using a strong teacher, such as LLM-based embedding models, and designs a teaching process in which various distance-based losses are used to minimise the distance between the embeddings produced by the student model and its teachers. The resulting models, the Stella and Jasper embedding families, are extremely strong embedding models at their respective sizes, and have been frequently demonstrated to reach very strong out-of-domain performance~\cite{rteb}.

As part of our training, we initially employed the partial codebase released by the Stella authors\footnote{\url{https://github.com/NovaSearch-Team/RAG-Retrieval}}, but found the full multi-step process difficult produce, yielding poorly performing models, with fluctuating performance and extremely high sensitivity to hyperparameters. Following work such as LEAF~\cite{leaf}, we opted to simplify the distillation loss to a simple L2 loss:
\begin{equation}
\mathcal{L}_2(y_i, \hat{y}_i) = \sum_{j=1}^{d} (y_{ij} - \hat{y}_{ij})^2
\end{equation}
which attempts to minimize the distance between our student's vectors and the teacher vectors.

As this step relies purely on embedding space distillation, there is no need to leverage retrieval datasets, since the relationship between queries and documents is not used during this stage. However, it has previously been highlighted that having a variety of inputs corresponding to common retrieval uses~\cite{leaf}, especially in terms of input lengths (e.g. longer documents and short queries), is helpful to improve performance. We thus sample many documents from various mixed sources and queries from large retrieval datasets, with a detailed data mix provided in Appendix~\ref{app:stelladata}.

We used StellaV5 1.5B as our teacher model, with an output dimension of 1024. We found that using higher teacher dimension embeddings resulted in decreasing performance, likely due to the vast dimension difference between our student models and the target sizes, while lower dimensions such as 768 resulted in mildly diminished results. To ensure our models' dimensions matched the teacher ones for distillation, we employed a 2-layer feedforward projection with a SiLU~\cite{silu}(a.k.a Swish~\cite{swish}) activation.

\begin{table}[h!]
\centering
\caption{Average NDCG@10 on NanoBEIR for dense Mxbai Edge variants.}
\begin{tabular}{lc}
\toprule
\textbf{Model} & \textbf{Avg. NDCG@10} \\
\midrule
Mxbai Edge 17M (Dense, FT) & 0.556 \\
Mxbai Edge 17M (Dense, Distill) & \textbf{0.567} \\
\midrule
Mxbai Edge 32M (Dense, FT) & 0.576 \\
Mxbai Edge 32M (Dense, Distill) & \textbf{0.626} \\
\bottomrule
\end{tabular}
\label{tab:distilresults}
\end{table}

Table~\ref{tab:distilresults} shows the NanoBEIR NDCG@10 results between the post-finetuning and post-distillation variants of both model sizes. It shows that, despite our streamlined process, this step results in performance gains\footnote{with the 32M variant reaching performance that is competitive with many state-of-the-art small embedding models.}, but unevenly distributed across model sizes. While the 32M variant heavily benefits from this step, the gains on the 17M model are more modest. We theorise that this might be due to the streamlined distillation loss we used struggling to bridge large dimensionality gaps compared to the original, more complex Stella loss mix, but do not explore this effect further.

\section{ColBERT Training}

Finally, we apply our final training stage for the ColBERT models. We run a series of ablations in order to create a strong baseline with this model, which will be able to support both real-world edge use cases and subsequent research uses satisfyingly.

We detail our training setting in the subsequent sections about our ablation work. For training data, we restrict ourselves to MSMARCO, so as to ensure that better data does not obscure the impact of training modifications (further details in Section~\ref{sec:data}), and use 16-way training tuples, where each query is associated with a positive example and 15 negative examples, all with a teacher score. We use a batch size of 128 and KL-Div loss with normalized scores~\cite{jacolbertv2.5}, except when otherwise specified. All experiments are performed using the PyLate~\cite{pylate} framework.

\subsection{Data}
\label{sec:data}

We experimented with various training datasets, comparing the MSMARCO \cite{msmarco} RLHN \cite{rlhn} set scored with Qwen3-Reranker~\cite{qwen3embed} as a teacher to the triplets used by answerai-colbert-small~\cite{answeraicolbert} and GTE-ModernColBERT~\cite{gtemoderncolbert}, with scores generated by BGE-Gemma2 Reranker\footnote{\url{https://huggingface.co/BAAI/bge-reranker-v2-gemma}}~\cite{bgegemma} to score a small subset of MSMARCO training tuples and comparing min-max normalized and unnormalized teacher scores~\cite{jacolbertv2.5}. 

\begin{table}[!ht]
\centering
\caption{Effect of teachers used for distillation.}
\setlength{\tabcolsep}{12pt}
\renewcommand{\arraystretch}{1.15}
\begin{tabular}{l c}
\hline
\textbf{Teacher} & \textbf{NDCG@10} \\
\hline
Qwen3\texttt{-}8B (no norm)      & 0.5991 \\
Qwen3\texttt{-}8B (minmax norm)  & 0.5854 \\
\textbf{BGE\texttt{-}Gemma2}         & \textbf{0.6286} \\
\hline
\end{tabular}
\label{tab:dataresults}
\end{table}

We present a comparison of these training methods in Table~\ref{tab:dataresults}. Surprisingly, whether normalized or unnormalized, Qwen3-Reranker is consistently outperformed as a teacher by the older BGE-Gemma2. Upon further analysis, it appears that on common retrieval benchmarks, the scores generated by Qwen3-Reranker are extremely skewed towards the extremes, with very scores outside of the [0.99,1] range for positives and [0, 0.01] range for negatives. We believe that this might be indicative of overfitting from the reranker, resulting in poor distributions to use for distillation. Our attempts at using both large and very small temperatures did not significantly change performance.

\subsection{Ablations}

We performed various ablations in order to understand the impact of certain parameters on model performance. To avoid overfitting, hyperparemeter ablations (optimizer, distillation impact, learning rate and projection dim) were evaluated on 5 NanoBEIR subset, so as to provide a good performance indicator without being exposed to the full BEIR sets. The selected subsets are high-quality search dataset MSMARCO (in-domain), SciFact (OOD), FiQA (OOD), NQ (OOD), and NFCorpus (OOD). Final ablations on projection layers and casing were evaluated on all of NanoBEIR.

\subsubsection{Optimizers}

We benchmarked both AdamW~\cite{adamw} and Muon~\cite{muon} across a range of learning rates with a fixed batch size. We present the results of these ablations in Table~\ref{tab:optims}. Our results indicate that even with limited experiments and the relatively small batch size that is commonly employed to train late-interaction models, Muon appears to be a strong optimizer for ColBERT model training.

\begin{table}[h]
\centering
\caption{Comparison of model performance across optimizers and learning rates. NDCG@10 is the average NDCG@10 score across the 4 ablation datasets.}
\begin{tabular}{l c}
\toprule
\textbf{Optimizer} & \textbf{NDCG@10} \\
\midrule
\textbf{AdamW} & \\
~1e-4 & 0.5911 \\
~5e-5 & 0.5780 \\
~8e-5 & 0.5923 \\
\midrule
\textbf{Muon} & \\
~1e-4 (AdamW 8e-5) & 0.5718 \\
~3e-4 (AdamW 8e-5) & 0.5604 \\
~5e-4 (AdamW 8e-5) & 0.5862 \\
~1e-3 (AdamW 8e-5) & \textbf{0.5985} \\
~3e-3 (AdamW 8e-5) & 0.5748 \\
\bottomrule
\end{tabular}
\label{tab:optims}
\end{table}

\subsubsection{Impact of Stella-style Distillation}

In Table~\ref{tab:stellacolbert}, we show a comparison of running trainings on the dense embedding result model resulting from our fine-tuning stage vs the model resulting from our distillation stage. Our results clearly show that Stella-style distillation improves performance of the resulting ColBERT model, even when projection heads are discarded to only retain the backbone model.

\begin{table}[h]
\centering
\caption{ColBERT performs better on a base embedding model trained on Stella-style distillation.}
\label{tab:stellacolbert}
\begin{tabular}{lc}
\toprule
\textbf{Base Model Variant} & \textbf{NDCG@10} \\
\midrule
32M model (fine-tuned only) & 0.5771 \\
32M model (with Stella-style distillation) & \textbf{0.5911} \\
\bottomrule
\end{tabular}
\end{table}

\subsubsection{Projection Dimension}

The projection dimension used by ColBERT models is traditionally set to 128, after the one used by the original ColBERT and ColBERTv2~\cite{colbert,colbertv2} models, and this dimension has shown good performance in both text and multimodal settings~\cite{colpali}. As of right now, the current state-of-the-art for smaller ColBERT models uses a projection dimension of 96~\cite{answeraicolbert}. In effect, the final projection dimension is largely defined in an arbitrary way, despite the large consequences it has in terms of both storage requirements and scoring speed.

\begin{table}[h]
\centering
\caption{Effect of projection dimension on NDCG@10 (Muon 1e-3, AdamW 8e-5) on the 32m model, using 20\% training data.}
\setlength{\tabcolsep}{14pt}
\renewcommand{\arraystretch}{1.2}
\begin{tabular}{c c}
\hline
\textbf{Projection Dimension} & \textbf{NDCG@10} \\
\hline
96 & \textbf{0.5991} \\
64 & 0.5985 \\
48 & 0.5967 \\
32 & 0.5772 \\
24 & 0.5423 \\
16 & 0.5126 \\
\hline
\end{tabular}
\label{tab:projectiondim}
\end{table}

In Table~\ref{tab:projectiondim}, we present the results of ablating a large range of projection dimensions, from 16 to 96. We show that lower dimensions hold up performance surprisingly well. Indeed, the performance decrease on NanoBEIR is very mild until a dimension of 48, but subsequently considerably degrades at projection dimensions of 32 and below.

\subsubsection{Projection Layers}

In a recent study, we demonstrated that the use of more complex projection layers outperformed the single-layer linear projection that is ubiquitous in ColBERT models~\cite{colbertprojections}. As part of this work, we experiment with the best variant proposed in our previous study, using a 2-layer feedforward network with an upscaled intermediate dimension and a residual connection, and compare it to a model trained with the ``normal'' ColBERT projection. 

\begin{table}[!ht]
\centering
\caption{Performance of different projection heads on the 17m model, under matched training hyperparameters, on full data.}
\setlength{\tabcolsep}{10pt}
\renewcommand{\arraystretch}{1.15}
\footnotesize
\begin{tabular}{lc}
\hline
\textbf{Projection} & \textbf{NDCG@10} \\
\hline
2-layer FFN & \textbf{0.6405} \\
Linear Projection  & 0.6275 \\
\hline
\end{tabular}
\label{tab:comparison}
\end{table}

We present the results of this comparison on the 17m parameter model variant in Table~\ref{tab:comparison}. As this experiment came later in our training process, results are reported as full NanoBEIR NDCG@10 rather the previously defined ablation sets. Our results that the use of better projection layers contributes positively to performance. While we do not perform significance testing due to the low number of evaluated checkpoints, we note that we reproduced this effect across a range of training seeds, with no single-layer linear projection checkpoint coming within less than 1 NDCG@10 of the 2-layer projection checkpoints.

\subsubsection{Casing}

Virtually all embedding models we consider "previous-generation" either use \textsc{bert-base-uncased}~\cite{bert} as their backbone, or models which were largely inspired by it. These encoders are all case-insensitive, meaning that all input text is lower-cased before being tokenized.

On the other hand, virtually all Large Language Models employ some form of casing, which ModernBERT~\cite{modernbert}, and thus subsequently Ettin~\cite{ettin}, also adopts.

The impact this tokenization change has, if any, has not yet been studied in detail. We decided to conduct an ablation in this sense, for which we present the results in Table~\ref{tab:lowercase}.

\begin{table}[!ht]
\centering
\caption{NanoBEIR NDCG@10 comparison with and without lower-casing as a pre-processing step, with all other hyperparameters kept equal and training on the full data.}
\begin{tabular}{l c}
\toprule
\textbf{Optimizer} & \textbf{NDCG@10} \\
\midrule
\textbf{32M} & \\
Lower-casing & \textbf{0.6520} \\
No lower-casing & \textbf{0.6519} \\
\midrule
\textbf{17M} & \\
Lower-casing & \textbf{0.6405} \\
No lower-casing & 0.6317 \\
\bottomrule
\end{tabular}
\label{tab:lowercase}
\end{table}

Our results demonstrate an interesting phenomenon: while there appears to be no significant difference at the 32M parameter scale, the results of the 17M model variant are significantly improved by lower-casing. Across random seeds, we also observed that lower-casing consistently reached stronger performance on the 17M model, but no discernible trend emerged at the 32M scale.

As with other ablations, we do not further attempt to understand the underlying mechanism, but theorise that the limited embedding dimensions and parameter count of the 17M model means that it benefits disproportionately from the learning simplification that lower-casing provides.

\section{Results}
\begin{table*}[h]
\caption{BEIR benchmark results (NDCG@10). Columns show the BEIR average and sampled tasks: MSMARCO, SciFact, Touche2020, FiQA, TREC-COVID, NQ, and DBPedia. Results in bold indicate best result for the weight class. The best results for size class are in \textbf{bold}. The complete table in Appendix~\ref{app:beirfull}.}
\begin{adjustwidth}{-3.2cm}{-0.1cm}
\centering

\setlength{\tabcolsep}{8pt}
\renewcommand{\arraystretch}{1.15}
\footnotesize
\begin{tabular}{lcccccccc}
\hline
\textbf{Model} & \textbf{AVG} & \textbf{MSMARCO} & \textbf{SF} & \textbf{Touche} & \textbf{FiQA} & \textbf{COVID} & \textbf{NQ} & \textbf{DBP} \\
\hline
\textbf{>100M parameters} & & & & & & & & \\
\midrule
GTE-ModernColBERT-v1 & \textbf{0.547} & 0.453 & \textbf{0.763} & \textbf{0.312} & \textbf{0.453} & \textbf{0.836} & \textbf{0.618} & \textbf{0.480} \\
ColBERTv2 & 0.488 & \textbf{0.456} & 0.693 & 0.263 & 0.356 & 0.733 & 0.562 & 0.446 \\
\midrule
\textbf{<35M parameters} & & & & & & & & \\
\midrule
\textbf{mxbai-edge-colbert-v0-32m} & 0.521 & \textbf{0.450} & \textbf{0.740} & \textbf{0.313} & 0.390 & 0.775 & \textbf{0.600} & 0.455 \\
answerai-colbert-small-v1 & \textbf{0.534} & 0.434 & \textbf{0.740} & 0.250 & \textbf{0.410} & \textbf{0.831} & 0.594 & \textbf{0.464} \\
bge-small-en-v1.5 & 0.517 & 0.408 & 0.713 & 0.260 & 0.403 & 0.759 & 0.502 & 0.400 \\
snowflake-s & 0.520 & 0.402 & 0.722 & 0.235 & 0.407 & 0.801 & 0.509 & 0.410 \\
\midrule
\textbf{<25M parameters} & & & & & & & & \\
\midrule
\textbf{mxbai-edge-colbert-v0-17m} & \textbf{0.490} & \textbf{0.416} & \textbf{0.719} & \textbf{0.316} & 0.326 & \textbf{0.713} & \textbf{0.551} & \textbf{0.410} \\
colbert-muvera-micro & 0.394 & 0.364 & 0.662 & 0.251 & 0.254 & 0.561 & 0.386 & 0.332 \\
all-MiniLM-L6-v2 & 0.419 & 0.365 & 0.645 & 0.169 & \textbf{0.369} & 0.472 & 0.439 & 0.323 \\
\hline
\end{tabular}
\end{adjustwidth}
\label{tab:beir_results}
\end{table*}

In Table~\ref{tab:beir_results}, we present the result of our model on a selected range of BEIR~\cite{beir} datasets and their average, while Table~\ref{tab:loco_results}, we present the result of our model on LongEmbed task~\cite{lemb}.

On the BEIR datasets, we note that our models are overall strong performers. While outperformed on short-context average by the previous small-scale state-of-the-art, answerai-colbert-small, they reach strong performance across the board. Particularly noteworthy is that mxbai-edge-colbert-v0-17m, a 17 Million parameter model, outperforms the still-widely-used ColBERTv2, despite having less than 1/6th of the parameters and a projection dimension set to just 48, a third of ColBERTv2's 128. They do so with remarkable efficiency, especially as context length increases, thanks to their ModernBERT-based backbone.

\begin{table*}[h]
\caption{Detailed LongEmbed benchmark performance. Context length is set to 4k and 32k context variants for models supporting it. Otherwise, it is set to the model's maximum sequence length (8k for granite-embeddings and 512 for others). Best results for size class are in \textbf{bold}, best overall results are \underline{underlined}. Models with more parameters than their size class but added for completeness are in \textit{italics}.}

\begin{adjustwidth}{-3.2cm}{-0.1cm} 
\centering
\setlength{\tabcolsep}{8pt}
\renewcommand{\arraystretch}{1.15}
\footnotesize
\begin{tabular}{l cccccccc}
\hline
\textbf{Model} & \textbf{AVG} & \textbf{NarrQA} & \textbf{QMSum} & \textbf{Wiki} & \textbf{SummScr.} & \textbf{Needle} & \textbf{Passkey} \\
\hline
\textbf{>100M parameters} & & & & & & & \\
\midrule
\textbf{GTE-ModernColBERT-v1 (32k)} & \textbf{\underline{0.898}} & \textbf{\underline{0.780}} & \textbf{\underline{0.737}} & \textbf{\underline{0.999}} & \textbf{\underline{0.953}} & \textbf{\underline{0.950}} & \textbf{0.970} \\
\textbf{GTE-ModernColBERT-v1 (4k)}  & 0.809 & 0.530 & 0.528 & 0.931 & 0.947 & \textbf{\underline{0.950}} & \textbf{0.970} \\
\textit{granite-embedding-english-r2}\footnote{} & 0.656 & 0.479 & 0.416 & 0.859 & 0.937 & 0.430 & 0.818 \\
ColBERTv2 & 0.428 & 0.287 & 0.254 & 0.648 & 0.686 & 0.330 & 0.365 \\
\midrule
\textbf{<50M parameters} & & & & & & & \\
\midrule
\textbf{mxbai-edge-colbert-v0-32m (32k)} & \textbf{0.849} & \textbf{0.585} & \textbf{0.698} & \textbf{0.993} & \textbf{0.910} & \textbf{0.915} & \textbf{\underline{0.990}} \\
\textbf{mxbai-edge-colbert-v0-32m (4k)} & 0.783 & 0.444 & 0.508 & 0.930 & 0.909 & \textbf{0.915} & \textbf{\underline{0.990}} \\
\textit{granite-embedding-small-english-r2}\footnote{}& 0.637 & 0.413 & 0.365 & 0.799 & 0.899 & 0.550 & 0.798 \\
answerai-colbert-small-v1 & 0.441 & 0.266 & 0.272 & 0.645 & 0.735 & 0.338 & 0.388 \\
bge-small-en-v1.5 & 0.312 & 0.220 & 0.208 & 0.430 & 0.532 & 0.263 & 0.218 \\
snowflake-arctic-embed-s & 0.356 & 0.177 & 0.230 & 0.411 & 0.643 & 0.283 & 0.390 \\
\midrule
\textbf{<25M parameters} & & & & & & & \\
\midrule
\textbf{mxbai-edge-colbert-v0-17m (32k)} & \textbf{0.847} & \textbf{0.621} & \textbf{0.733} & \textbf{0.977} & \textbf{0.943} & \textbf{\underline{0.950}} & \textbf{0.858} \\
\textbf{mxbai-edge-colbert-v0-17m (4k)} & 0.776 & 0.437 & 0.566 & 0.909 & 0.935 & \textbf{\underline{0.950}} & \textbf{0.858} \\
all-MiniLM-L6-v2 & 0.298 & 0.183 & 0.163 & 0.463 & 0.548 & 0.200 & 0.233 \\
colbert-muvera-micro & 0.405 & 0.230 & 0.244 & 0.566 & 0.689 & 0.318 & 0.385 \\
\hline
\end{tabular}
\end{adjustwidth}
\label{tab:loco_results}
\end{table*}
\footnotetext[9]{149M parameter model. Results are from the MTEB leaderboard.}
\footnotetext[10]{49M parameter model. Results are from the MTEB leaderboard.}

On long-context evaluations, our models reach very strong performance, only outperformed again by the larger GTE-ModernColBERT. We show that both of our models are extremely strong performer, only outperformed by the larger GTE-ModernColBERT. As expected, models based on more modern architecture, capable of handling longer context lengths, are the only competitive models on this task, with previous methods unable to process longer documents efficiently and resorting to truncation, thus greatly reducing their performance.

Particularly notably, even our 17M parameter variant outperforms the current <1B parameter single-vector retrieval state-of-the-art\footnote{According to the MTEB Leaderboard as of October 2025} on LongEmbed tasks, such as granite-embedding-r2, by almost 20NDCG@10 points. Note that Needle and Passkey are computed on NDCG@1 and are calculated by taking the average of all lengths.

 Interestingly, we note, similarly to \cite{gtemoderncolbert} that despite being based on a model with a native 8,000 context window, mxbai-edge-colbert's models are capable of handling 32k sequence lengths and observe performance gains from it, despite our retrieval training using documents truncated to 220 tokens.

Finally, we note that the low parameter count of our models, in combination with their highly-efficient architecture, make them particularly suitable for reranking task. This is especially true for longer chunks, as there currently does not exist any re-ranker able to reach similarly strong performance while running with low latencies on CPU for long document reranking.

\subsubsection{Efficiency comparison with other small ColBERTs}

\begin{table*}[ht!]
\centering
\caption{Relative performance and efficiency comparisons of small ColBERT models on NanoBEIR, with ColBERTv2 as a reference. CPU and GPU refer to runtimes as the average of 10 runs on each hardware type. Mem. is RAM requirements, in MB, for storing 10,000 300 token document representations in fp16. LoCo. stands for long-context support. Dim. is the projection dimension of each model. Best values are in \textbf{bold}, best values while outperforming ColBERTv2 on retrieval are \underline{underlined}.}
\begin{tabular}{l|c|c|c|c|cc|c}
\toprule
\textbf{Model} & \textbf{Params} & \textbf{Dim.} & \textbf{NDCG@10} & \textbf{LoCo} & \textbf{GPU} & \textbf{CPU} & \textbf{Mem.} \\
\midrule
ColBERTv2 & 130M & 128  & 0.6198 & -- & 81s & 1540s & 732 \\
\midrule
answerai-colbert-small-v1 & 33M & 96 & \underline{\textbf{0.6545}} & -- & 59s & 621s & 549 \\
colbert-muvera-micro & 4M & 128 & 0.5599 & -- & \textbf{45s} & \textbf{88s}  & 732 \\
\midrule
mxbai-edge-colbert-v0-17m & 17M & 48 & 0.6405 & \checkmark & \underline{51s} & \underline{487s} &  \underline{\textbf{275}} \\
mxbai-edge-colbert-v0-32m & 32M & 64 & 0.6520 & \checkmark & 55s & 589s & 366 
\\
\bottomrule
\end{tabular}
\label{nanoedge}
\end{table*}

Table~\ref{nanoedge} shows the relative performance of our 17M parameter edge ColBERT against other commonly used small ColBERTs, along with efficiency comparisons. For ease of rapid evaluation, we report overall NanoBEIR NDCG@10 scores, long-context support, projection dimension (a very important factor for edge use cases), mean runtime of 10 NanoBEIR evaluation runs on both GPU with a single RTX 4090 and CPU, representing the encoding of around 67,000 documents, as well as 650 queries and as many searches and scoring steps. 

We also report the memory usage required to store the 16-bit vector representations of 10,000 300-tokens documents stored, a direct factor of the projection dimension, to provide a brief overview of the suitability for various in-memory encoding usages.

\section{Conclusion}

We introduce v0 of the Mxbai-edge-ColBERT model family. These models represent the first small ColBERT model to fully benefit from a modern architecture, with long-context support and all the efficiency improvements introduced by the ModernBERT~\cite{modernbert} generation of encoder models.

Our intent with these models is two-fold: our main aim is for them to provide a suitable testbed for future experiments and distillation of our research on larger-scale models, as well as to serve as a strong performance predictor experiments, following scaling laws. Our second aim to support a large range of on-device use-cases, be it local RAG projects or extremely efficient re-ranking on both CPU and GPU.

Mxbai-edge-ColBERT-v0, at both model sizes, reach strong performance on a variety of datasets. Notably, the 17M parameter variant outperforms ColBERTv2 with an order-of-magnitude fewer parameters, and with vector storage and scoring-time compute requirements reduced by two thirds.

We fully intend to continue to upgrade these models with future developments and are looking forward to seeing them used in the real-world.

\bibliographystyle{splncs04}
\bibliography{bib}

\appendix

\section{Distillation Data}
\label{app:stelladata}

The data mix for the distillation stage is provided in Tables~\ref{tab:query} and \ref{tab:passage}.

\begin{table}[!ht]
\centering
\setlength{\tabcolsep}{10pt}
\renewcommand{\arraystretch}{1.1}
\footnotesize
\begin{minipage}[h]{0.48\textwidth}
\centering
\caption{Queries used for distillation}
\begin{tabular}{l r}
\hline
\textbf{Dataset} & \textbf{Size (rows)} \\
\hline
msmarco & 510k \\
amazon\_qa & 475k \\
nq & 175k \\
triviaqa & 70k \\
pubmed & 67k \\
arxiv & 50k \\
cornstk & 50k \\
lotte & 25k \\
medqa & 13k \\
mldr & 12.2k \\
\hline
\textbf{Total} & \textbf{1.45M} \\
\hline
\end{tabular}
\label{tab:query}
\end{minipage}\hfill
\begin{minipage}[h]{0.48\textwidth}
\centering
\caption{Passages used for distillation}
\begin{tabular}{l r}
\hline
\textbf{Dataset} & \textbf{Size (rows)} \\
\hline
DCLM-Pro & 1.59M \\
english\_words & 742k \\
fineweb & 665k \\
dclm\_sent & 400k \\
ccnews & 370k \\
stack & 185k \\
ettin\_tokens & 50k \\
\hline
\textbf{Total} & \textbf{4.00M} \\
\hline
\end{tabular}
\label{tab:passage}
\end{minipage}
\end{table}

DCLM-Pro~\cite{dclmpro} and FineWeb~\cite{fineweb} are full documents randomly from their respective datasets, while dclm\_sent is comprised of individual DCLM-Pro documents broken down into individual sentences, to create more varied small-length inputs, again following Stella~\cite{stella}. ettin\_tokens and english\_words were added during the course of this study following the release of LEAF~\cite{leaf}, which used a similar method to improve trianing. ettin\_tokens is a dataset comprised of very short input, where each document is a single token from our model's tokenizer, while english\_words is a large collection of English words along with a definition generated by Gemini 2.0 Flash.

\section{Full BEIR Results}
\label{app:beirfull}
We show the full BEIR results in Tables~\ref{tab:beir_full_part_a} and \ref{tab:beir_full_part_b}.
\begin{table*}[h]
\centering
\caption{BEIR benchmark (Part A): AVG and (Touche2020, NQ, MSMARCO, SciFact, FiQA2018, NFCorpus, ArguAna). Scores are NDCG@10.}
\setlength{\tabcolsep}{5.5pt}
\renewcommand{\arraystretch}{1.08}
\scriptsize
\noindent\hspace*{-0.3\textwidth}
\resizebox{1.6\textwidth}{!}{%
\begin{tabular}{lcccccccc}
\hline
\textbf{Model} & \textbf{AVG} & \textbf{Touche2020} & \textbf{NQ} & \textbf{MSMARCO} & \textbf{SciFact} & \textbf{FiQA2018} & \textbf{NFCorpus} & \textbf{ArguAna} \\
\hline
\textbf{>100M parameters} & & & & & & & & \\
\hline
GTE-ModernColBERT-v1 & \textbf{\underline{0.547}} & \textbf{0.312} & \textbf{\underline{0.618}} & 0.453 & \textbf{\underline{0.763}} & \textbf{\underline{0.453}} & \textbf{\underline{0.379}} & \textbf{\underline{0.485}} \\
colbertv2             & 0.488 & 0.263 & 0.562 & \textbf{\underline{0.456}} & 0.693 & 0.356 & 0.338 & 0.463 \\
\hline
\textbf{<35M parameters} & & & & & & & & \\
\hline
mxbai-edge-colbert-v0-32m & 0.521 & \textbf{0.313} & \textbf{0.600} & \textbf{0.450} & \textbf{0.740} & 0.390 & 0.362 & 0.454 \\
answerai-colbert-small-v1 & \textbf{0.533} & 0.250 & 0.594 & 0.434 & \textbf{0.740} & \textbf{0.410} & \textbf{0.369} & \textbf{0.468} \\
bge-small-en-v1.5 & 0.517 & 0.260 & 0.502 & 0.408 & 0.713 & 0.403 & 0.349 & 0.331 \\
snowflake-s & 0.520 & 0.235 & 0.509 & 0.402 & 0.722 & 0.407 & 0.324 & 0.339 \\
\hline
\textbf{<25M parameters} & & & & & & & & \\
\hline
mxbai-edge-colbert-v0-17m & \textbf{0.490} & \textbf{\underline{0.316}} & \textbf{0.551} & \textbf{0.416} & \textbf{0.719} & 0.326 & \textbf{0.352} & \textbf{0.464} \\
colbert-muvera-micro      & 0.394 & 0.251 & 0.386 & 0.364 & 0.662 & 0.254 & 0.321 & 0.303 \\
all-MiniLM-L6-v2          & 0.419 & 0.169 & 0.439 & 0.365 & 0.645 & \textbf{0.369} & 0.314 & 0.331 \\
\hline
\end{tabular}%
}
\label{tab:beir_full_part_a}
\end{table*}

\begin{table*}[h]
\centering
\caption{BEIR benchmark (Part B): Rest of the tasks (QuoraRetrieval, SCIDOCS, TRECCOVID, ClimateFEVER, HotpotQA, DBPedia, CQADupstack, FEVER). Scores are NDCG@10.}
\setlength{\tabcolsep}{5.5pt}
\renewcommand{\arraystretch}{1.08}
\scriptsize
\noindent\hspace*{-0.3\textwidth}
\resizebox{1.6\textwidth}{!}{%
\begin{tabular}{lcccccccc}
\hline
\textbf{Model} & \textbf{QuoraRetrieval} & \textbf{SCIDOCS} & \textbf{TRECCOVID} & \textbf{ClimateFEVER} & \textbf{HotpotQA} & \textbf{DBPedia} & \textbf{CQADupstack} & \textbf{FEVER} \\
\hline
\textbf{>100M parameters} & & & & & & & & \\
\hline
GTE-ModernColBERT-v1 & \textbf{0.866} & \textbf{0.191} & \textbf{\underline{0.836}} & \textbf{0.306} & \textbf{\underline{0.773}} & \textbf{\underline{0.480}} & \textbf{0.410} & \textbf{0.874} \\
colbertv2             & 0.852 & 0.154 & 0.733 & 0.176 & 0.667 & 0.446 & 0.378 & 0.785 \\
\hline
\textbf{<35M parameters} & & & & & & & & \\
\hline
mxbai-edge-colbert-v0-32m & 0.863 & 0.170 & 0.775 & 0.290 & 0.734 & 0.455 & 0.388 & 0.826 \\
answerai-colbert-small-v1 & 0.879 & 0.187 & \textbf{0.831} & \textbf{\underline{0.328}} & \textbf{0.769} & \textbf{0.464} & 0.394 & \textbf{\underline{0.887}} \\
bge-small-en-v1.5 & \textbf{\underline{0.887}} & \textbf{0.198} & 0.759 & 0.253 & 0.699 & 0.400 & 0.391 & 0.866 \\
snowflake-s  & 0.884 & 0.218 & 0.801 & 0.352 & 0.665 & 0.410 & \textbf{0.397} & 0.871 \\
\hline
\textbf{<25M parameters} & & & & & & & & \\
\hline
mxbai-edge-colbert-v0-17m & 0.839 & 0.169 & \textbf{0.713} & \textbf{0.224} & \textbf{0.713} & \textbf{0.410} & 0.356 & \textbf{0.784} \\
colbert-muvera-micro      & 0.764 & 0.123 & 0.561 & 0.115 & 0.528 & 0.332 & 0.313 & 0.637 \\
all-MiniLM-L6-v2          & \textbf{0.876} & \textbf{\underline{0.217}} & 0.472 & 0.203 & 0.465 & 0.323 & \textbf{\underline{0.412}} & 0.519 \\
\hline
\end{tabular}%
}
\label{tab:beir_full_part_b}
\end{table*}

\end{document}